\documentclass{optica-article}
\journal{opticajournal} % for journals or Optica Open
\articletype{Research Article}
\usepackage{booktabs}
\usepackage{siunitx}
\DeclareSIUnit\bar{bar} % \bar is depreciated and needs to be self defined

\begin{document}

\title{Soft X-ray Reflection Ptychography}

\author{Damian Guenzing,\authormark{1,*} Dayne Y. Sasaki,\authormark{1,2,*} Alexander S. Ditter,\authormark{1} Abraham L. Levitan,\authormark{3} Eric M. Gullikson,\authormark{4} Scott Dhuey,\authormark{5} Arian Gashi,\authormark{5} Hendrik Ohldag,\authormark{1} Sujoy Roy,\authormark{1} David A. Shapiro,\authormark{1} Riccardo Comin,\authormark{2, +} and Sophie A. Morley\authormark{1, +}}

\address{\authormark{1}Advanced Light Source, Lawrence Berkeley National Laboratory, Berkeley, CA, USA\\
\authormark{2}Department of Physics, Massachusetts Institute of Technology, Cambridge, MA, USA\\
\authormark{3}Center for Photon Science, Paul Scherrer Institute PSI, Villigen, Switzerland\\
\authormark{4}Center for X-Ray Optics, Lawrence Berkeley National Laboratory, Berkeley, CA, USA\\
\authormark{5}Molecular Foundry, Lawrence Berkeley National Laboratory, Berkeley, CA, USA}

\email{\authormark{+}Correspondence to: smorley@lbl.gov, rcomin@mit.edu}
\noindent\authormark{*}These authors contributed equally to this work.

% use {asbstract*} to suppress the copyright line. Copyright information will be added in production

\begin{abstract*}
Scanning transmission X-ray microscopy and ptychography have become mature tools for high-resolution, element-specific imaging of nanoscale structures. However, transmission geometries impose stringent constraints on sample thickness and preparation, thereby limiting investigations of extended or bulk specimens, especially in the soft X-ray region. Here, we demonstrate reflection‑geometry soft X-ray ptychography as a robust imaging mode. Instrumental feasibility and spatial resolution are established using a lithographically defined Siemens star and barcode test pattern on a multilayer substrate. We empirically demonstrate a full-pitch spatial resolution of ca. \SI{45}{\nano\metre} from Fourier ring correlation analysis of the reconstructed object. The results highlight the potential of the reflection geometry for nondestructive X-ray studies of materials without the need for transmissive samples.
\end{abstract*}

%%%%%%%%%%%%%%%%%%%%%%%%%%  body  %%%%%%%%%%%%%%%%%%%%%%%%%%

\section{Introduction}
X-ray microscopy has revolutionized the way we investigate the chemical, geometric, electronic and magnetic structure of matter on the nanoscale. However, current state-of-the-art soft X-ray ptychographic microscopes almost exclusively operate in a transmission geometry: the specimen is illuminated by a focused, (partially) coherent beam and the transmitted photons are detected behind the sample. Transmission setups benefit from relatively simple optics and straightforward data interpretation, and they form the basis of widely used techniques such as scanning transmission X-ray microscopy (STXM) \cite{kilcoyne2003, gianoncelli2016}, transmission X-ray microscopy (TXM) \cite{andrews2011} and coherent diffraction imaging (CDI) \cite{giewekemeyer2011, salditt2020}. Combined with X-ray selectivity at elemental resonances and polarized X-rays, magnetic and electronic information can be obtained \cite{rahmim2002, tripathi2011, shapiro2020, butcher2024, butcher2025}. However, the requirement for the sample thickness to be within the X-ray penetration depth (typically a few tens to hundreds of nanometers in the soft X-ray regime) imposes severe limitations on the range of accessible materials. One possibility to push the boundaries of measurable sample thicknesses is to utilize pre-edge magnetic phase contrast in the soft X-ray region as recently shown for thicker magnetic samples \cite{neethirajan2024}. However, bulk crystals, multilayer stacks, integrated devices, or any sample grown on an X-ray opaque substrates are effectively incompatible with transmission-based microscopy instruments.

An X-ray imaging system capable of operating in a reflection geometry could overcome these limitations of traditional transmission-based microscopes. By probing the topmost sample depth at grazing incidence, chemical and magnetic information can be obtained without thinning, polishing, or otherwise perturbing the specimen. X-ray photoemission electron microscopy (XPEEM) enables element-specific reflection imaging at resolutions of tens of nanometers but requires conductive samples, the photoelectron emission and detection complicates in situ magnetic and electrical measurements, and it is limited to surface-sensitive detection within a few nanometers depth \cite{renault2007, nemsak2017, locatelli2025}. A similar recently demonstrated surface-sensitive technique, direct coherent X-ray imaging, has been applied to study antiferromagnetic textures in MnBi$_2$Te$_4$ making use of antiphase domain boundary contrast \cite{burgard2025}. There are also limited examples of coherence-based reflection imaging which uses holography and requires a reference and imaging pinhole screen mounted a few tens of micrometers away from the sample surface, making it technically challenging and preventing its widespread adoption \cite{roy2011, popescu2021}. 

Reflection-mode ptychography on the other hand offers an attractive avenue towards enabling high-resolution imaging capabilities. Ptychography is a technique which raster scans the sample (object) with a coherent illumination (probe) to collect a series of coherent diffraction patterns from overlapping beam positions. Computational phase retrieval algorithms are then used to reconstruct the complex-valued images of the object and probe from the collected data. 

Experimental studies of reflection-mode ptychography have demonstrated resolutions of \SIrange{50}{150}{\nano\metre} \cite{seaberg2014, tanksalvala2021, shao2024} in the EUV range while ca. \SIrange{20}{40}{\nano\metre} spatial resolution is possible in the hard X-ray region albeit with highly anisotropic resolution due to the elongated beam footprint from the ultra low incident grazing angles \cite{burdet2017, jorgensen2024}. Nevertheless combining grazing incidence ptychography and reflectivity can yield unique information about the electronic structure and thickness simultaneously \cite{myint2024}.  Additionally, Bragg ptychography can be utilized to deduce magnetic and ferroelectric properties, as well as crystal strain information in a sample \cite{godard2011, hruszkewycz2012, hruszkewycz2013, hruszkewycz2016, Longlong2021}. However, this technique can only be applied to samples hosting an underlying crystal lattice with an appropriate lattice spacing. Recently, the feasibility of soft x-ray reflection ptychography was reported, although only moderate image quality and resolution was achieved \cite{popescu2021}.  As transmission soft X-ray ptychography setups now demonstrate resolutions down to ca. \SI{8}{\nano\metre} \cite{shapiro2020}, a potential opportunity exists for corresponding reflection-mode microscopes to reach similar performance levels.

In this article we report on the design and characterization of a soft X-ray reflection ptychographic microscope, demonstrating the first high quality images for this technique with ca. $\SI{45}{\nano\metre}$ resolution. Following an overview of the instrumental concept, we benchmark the spatial resolution using a nano-fabricated Siemens star pattern. The work establishes reflection ptychography as a versatile probe for materials compatible with in situ electrical and magnetic fields, with the added potential for depth-sensitivity.

\section{Experimental Details and Setup}\label{sec:exp}
\begin{figure}[htbp]
    \centering 
    \includegraphics[width=0.6\textwidth]{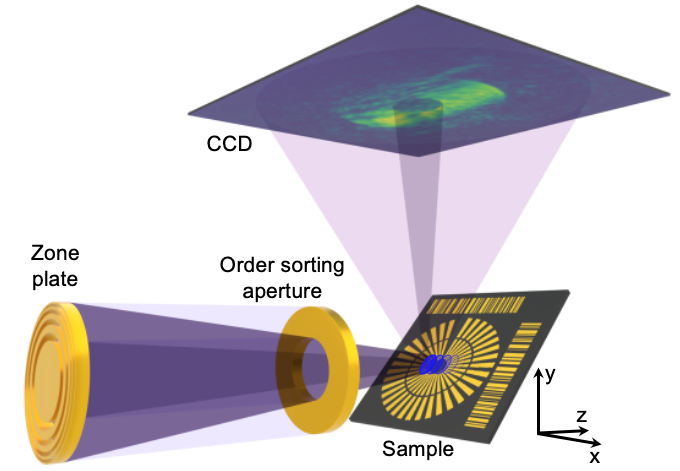} 
    \caption{Schematic of the experimental setup. A zone plate and order sorting aperture are used to illuminate the sample with a convergent beam of coherent soft x-rays. A ptychography dataset is collected by defocusing the zone plate illumination and using a charge coupled device (CCD) to measure coherent scattering patterns from overlapping beam positions on the sample as indicated with the transparent blue ellipses. The coherent scattering pattern intensity is shown at the CCD on a logarithmic scale.}
    \label{fig:instrument}
\end{figure}

Experiments were performed at the soft X-ray imaging branch of Coherent Scattering and Microscopy (COSMIC Imaging) beamline~7.0.1.2 at the Advanced Light Source (ALS). The X-ray source is an elliptically polarizing undulator delivering photons with variable linear or circular polarization in the energy range \SIrange{250}{2500}{\electronvolt}. A plane grating monochromator provides an energy resolving power $E/\Delta E$ of 2000 and a coherent flux at the zone plate of ca. $10^{11}$~photons~s$^{-1}$ at \SI{1}{\kilo\electronvolt}, while the sample receives ca. $10^{9}$~photons~s$^{-1}$ through the first-order focused light.

%The reflection microscopy setup (Fig.~\ref{fig:instrument}) is based on a zone--plate focusing optic with outermost zone width $\Delta r = \SI{0000}{\nano\metre}$ and diameter $D = \SI{0000}{\micro\metre}$. A central stop and order--sorting aperture (OSA) suppress unwanted diffraction orders. The incidence angle $\theta$ onto the sample surface can be adjusted between \SI{10}{\degree} (grazing) and \SI{90}{\degree} (normal), enabling a compromise between surface sensitivity and ease of alignment. The backscattered intensity is collected by a large--area photodiode mounted off the specular direction to minimise contributions from the direct reflection of the incident beam. All optical elements are mounted on piezo stages, providing \SI{<10}{\nano\metre} positioning accuracy within a vibration--damped vacuum chamber (base pressure \SI{1e-7}{\milli\bar}).

The reflection microscopy setup (Fig.~\ref{fig:instrument}) builds upon the Nanosurveyor STXM endstation at COSMIC Imaging (henceforth referred to as Nanosurveyor), which operates as both a conventional STXM and transmission X-ray ptychographic microscope. Details of this endstation are discussed elsewhere \cite{shapiro2020}; however we briefly review features of this instrument here for clarity. A zone plate focusing optic with outermost zone width $\Delta r = \SI{45}{\nano\metre}$ and diameter $D = \SI{360}{\micro\metre}$ was used, together with a central beam stop of $D = \SI{95}{\micro\metre}$ and an order-sorting aperture (OSA) to suppress unwanted diffraction orders. The OSA was placed ca. $\SI{10.5}{\milli\metre}$ downstream of the zone plate, with the sample positioned ca. $\SI{2}{\milli\metre}$ downstream of the OSA. All optical elements are mounted on piezo stages, providing \SI{<10}{\nano\metre} positioning accuracy within a vibration-damped vacuum chamber (base pressure \SI{1e-7}{\milli\bar}). A charge-coupled device (CCD) detector (ANDOR iKon-L, Oxford Instruments) with 2048 $\times$ 2048 pixels and a $\SI{13.5}{\micro\metre}$ pixel size was mounted on Nanosurveyor at a fixed 2$\theta$ angle of \SI{90}{\degree} with a sample-to-detector distance of ca. $\SI{156}{\milli\metre}$.

To maximize the specularly-reflected X-ray flux at this right-angle scattering geometry a multilayer stack of [Si/W]$_{100}$ deposited on a Si substrate was used. The multilayer is comprised of a [Si/W] bilayer with thickness of ca. \SI{1.2}{\nano\metre} and 100 periods, optimized for enhanced specular reflection at photon energy, E = $\SI{707}{\electronvolt}$. The [Si/W] bilayer thickness was confirmed from X-ray reflectometry measurements shown in supplementary Fig. \ref{fig:SI_XRR}.  Compared to a bare Si substrate, this multilayered substrate (henceforth referred to as the multilayer) increased the reflectivity signal by ca. 3 orders of magnitude at 2$\theta$ = \SI{90}{\degree} . On top of this substrate, a \SI{3}{\nano\metre} Ti \ / \SI{32}{\nano\metre} Au pattern comprising of a Siemens star and barcodes was fabricated using electron-beam lithography and electron-beam evaporation. The scanning electron microscopy (SEM) image is shown in Fig.~\ref{fig:resolution} (a). The barcode lines in this pattern have a fixed length of \SI{1}{\micro\meter} and widths ranging from \SIrange{15}{200}{\nano\meter}. The vertically and horizontally oriented barcode patterns are identical and differ by a rotation of \ang{90}. The linewidths of the tips and ends of the Siemens star are ca. \SI{40}{\nano\metre} and ca. \SI{250}{\nano\metre}, respectively. The lengths of the inner and outer spokes of the Siemens star are both ca. \SI{1.2}{\micro\metre} and have a separation of ca. \SI{120}{\nano\metre}.

% ----- The following can be inserted if we want to talk more about near-field and/or alignment maybe SI ? 
% The sample and area of interest were located by using a highly defocused beam (approximately $\SI{1}{\micro\metre}$ beam spot), which produces an in-line hologram on the detector. This hologram appears as a full-field ``image" of the sample with features surrounded by fringes as shown in XXX. Once the area of interest has been centered on the hologram, the beam was first coarsely focused on the sample surface by visually maximizing the speckle size on the detector (speckle size grow inversely proportional to the beam size) by manually moving the zone plate in/out of focus. A finer focusing was performed by collecting 1D line scans across several bars as a function of zone plate distance and choosing the distance which produced the sharpest line scans.

Ptychography data is typically acquired at Nanosurveyor by scanning the optic within a plane perpendicular to the optical axis. For scanning reflection microscopy, motion along $y$ requires a tandem motion of the optic along $z$ to maintain a constant beam size across the sample surface. The ptychographic scans have been performed over a \SI{10}{\micro\meter} $\times$ \SI{10}{\micro\meter} area, with a $50 \times 50$ grid corresponding to a \SI{200}{\nano\meter} step size, an acquisition time of \SI{1500}{\milli\second} per position, and with the zone plate positioned ca. \SI{10.7}{\milli\metre} upstream of the sample, resulting in an approximately \SI{1.5}{\micro\meter} diameter beam spot at the sample. 

For phase retrieval in the reflection geometry, several approaches have been developed. Modified versions of the ptychographical iterative engine (PIE) \cite{maiden2009} have been adapted for reflection geometry, including angular auto-calibration in reflection ptychography (aPIE) \cite{beurs2022}, which can be implemented using software packages such as PtyLab \cite{loetgering2023}. Alternative reconstruction methods employ automatic differentiation (AD) \cite{kandel2019, senhorst2024}, which has also demonstrated efficacy for subpixel position corrections in ptychography \cite{xu2025}. The reflection ptychography dataset presented here was reconstructed using the \textit{CDTools} ptychography package employing a transmission geometry basis \cite{cdtools}, with its PyTorch-based Adam (Adaptive Moment Estimation) optimizer implementation. It has been shown that the Adam optimizer is a robust tool for ptychographic reconstructions in reflection geometry with EUV light \cite{senhorst2024}. Optimal reconstruction quality was achieved using one probe mode. The learning rate was decreased every 100 iterations by a factor of 10 to allow the model to make finer adjustments and converge more stably as training progressed.

\section{Results and Discussion}
\begin{figure}[tb]
\centering
\includegraphics[width=0.70\textwidth]{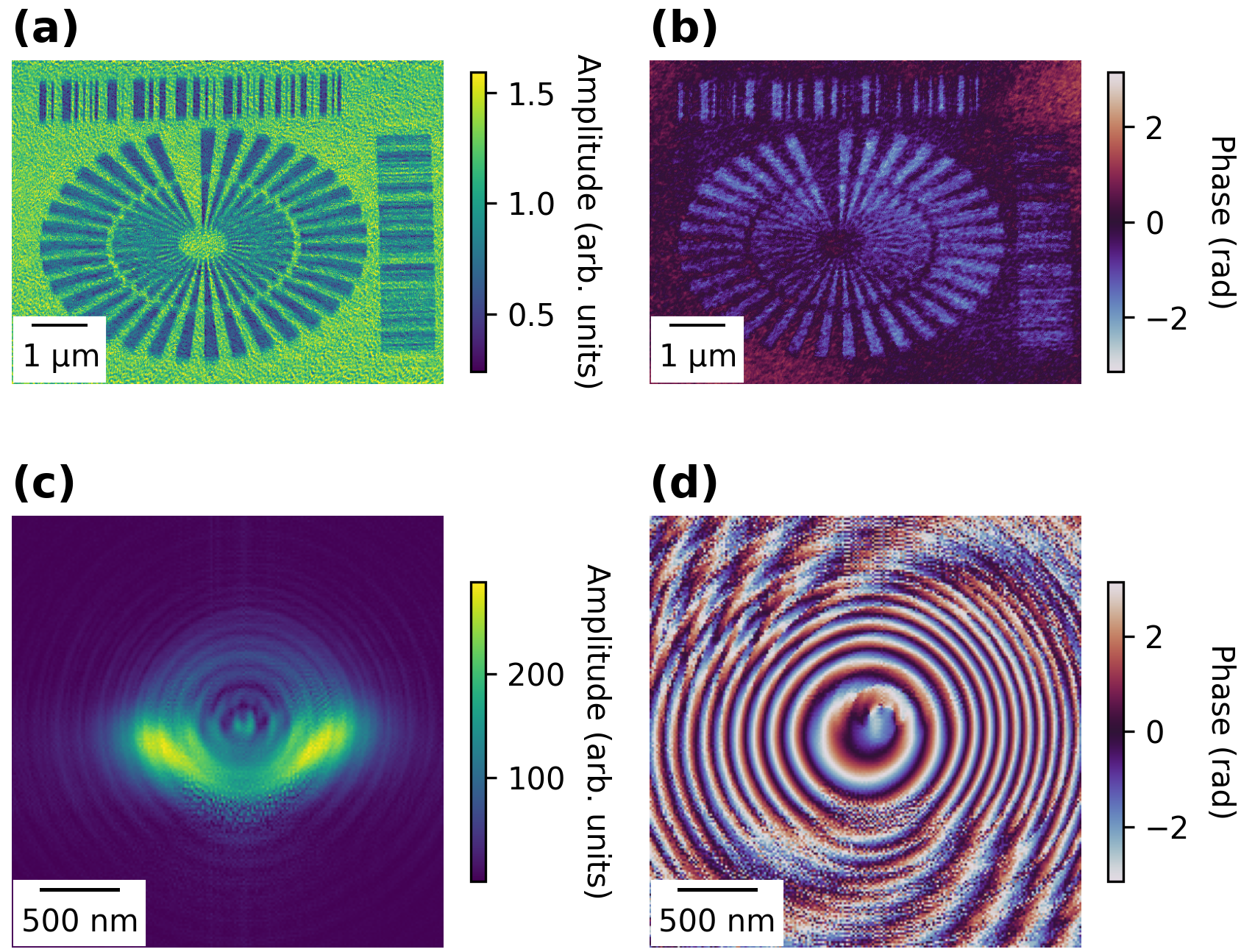}
\caption{Ptychographic reconstruction results obtained using \textit{CDTools}, (a) reconstructed object amplitude and (b) phase distributions and the (c) amplitude and (d) phase of the retrieved probe.}
\label{fig:reconstruction}
\end{figure}

The amplitude and phase of the reconstructed object are shown in Fig.~\ref{fig:reconstruction}~(a,b) along with the corresponding amplitude and phase of the retrieved probe in Fig.~\ref{fig:reconstruction}~(c,d) obtained through the ptychographic phase retrieval. For the barcode pattern near the top of the image, we are able to visualize all the patterned lines, including the finest line with widths down to ca. $\SI{15}{\nano\metre}$. The gold structures exhibit a lower amplitude because they absorb most of the X-rays. The phase shift is also clearly visible between the pure multilayer and the nano-patterned gold. The amplitude of the reconstructed probe arises from an interaction of the incoming illumination, produced by the Fresnel zone plate, with the multilayer. As will be discussed later, the multilayer angularly filters the incident zone plate illumination, which results in a prominent bright "band" appearing in both the retrieved probe and the bright field annuli of the diffraction patterns (see Fig.~\ref{fig:instrument} and supplementary Fig. \ref{fig:SIexamplepattern}). In the phase, the same higher amplitude region shows more structure.

\begin{figure}[tb] % 'h' means to place the figure here
    \centering\includegraphics[width=0.995\textwidth]{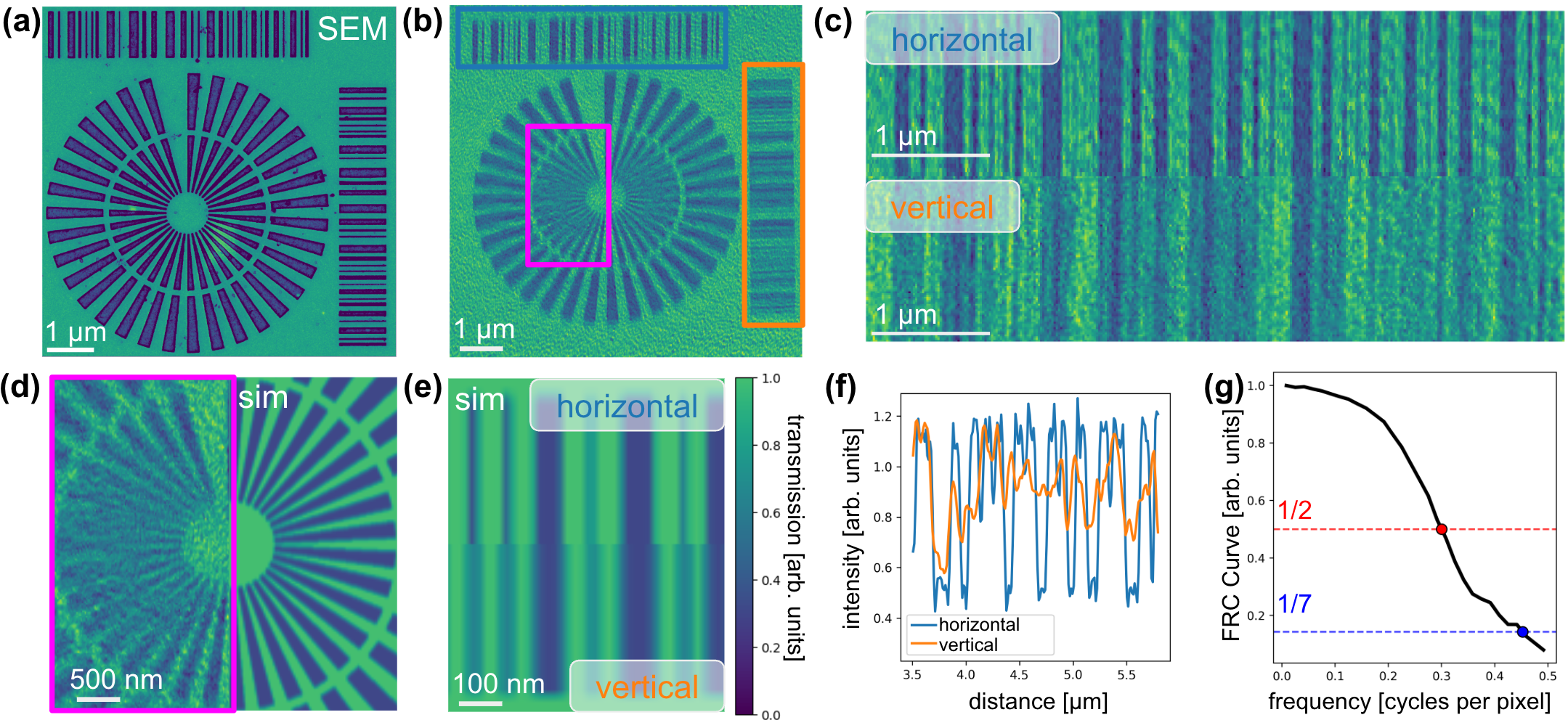} 
    \caption{(a) Scanning electron microscopy image of the Siemens star with a horizontal and vertical barcode, (b) reflection ptychography image where boxed area used for 1D line-cuts are marked for the horizontal (blue) and vertical (orange) directions. The beam propagation and tilt are along the vertical and horizontal is parallel to the rotation axis. (c) Comparison between the horizontal and vertical barcode pattern as indicated in (b). (d) Comparison of the reconstruction and simulation of the star center (e) simulated transmitted signal through the $\SI{32}{\nano\metre}$ of gold for the horizontal and vertical bar pattern, (f) line-cut plots through the barcode reconstruction, and (g) FRC curve with thresholds determining the resolution estimations.}
    \label{fig:resolution}
\end{figure}

% /// THE ORIGINAL DISCUSSION ABOUT VERTICAL COMPRESSION BEFORE DAYNE MERGED THE PARAGRAPHS
We observe several differences between the object amplitude as imaged from the reflection ptychography and the sample as imaged using an SEM. This comparison can be more easily seen when we plot them together in Fig.~\ref{fig:resolution}~(a) and (b), respectively. For instance, the actual Siemens star pattern has a circular shape (Fig.~\ref{fig:resolution}~(a)), whereas the reconstructed pattern has an elliptical shape and the pattern set as a whole appears vertically compressed. Moreover, an anisotropic blurring is observed, where the sharpness of the patterns depend on the extent of parallel alignment between the optical axis and the lengths of the Siemens star spokes and barcode lines. Parallel alignment between the feature lengths and the optical axis results in the sharpest features while perpendicular alignment produces the most blurred appearance. This behavior can be seen more clearly at the inner spokes of the Siemens star shown zoomed in Fig.~\ref{fig:resolution}~(d) (left) and also when directly comparing the reconstructed horizontal and vertical barcode patterns in Fig.~\ref{fig:resolution}~(c). To assess this further, we simulated an amplitude image of our test sample (right-half of Fig.~\ref{fig:resolution}~(d) and Fig.~\ref{fig:resolution}~(e)) by calculating the X-ray transmission through a three-dimensional model of the test sample at a \ang{45} incidence geometry, treating the multilayer substrate as an ideal mirror for simplicity (see supplementary \ref{sec:simulation}). There is relatively good agreement between our simulation and the anisotropic blurring observed experimentally.

% /// THE ORIGINAL DISCUSSION ABOUT ANISOTROPIC BLURRING BEFORE DAYNE MERGED THE PARAGRAPHS
The vertical compression and anisotropic blurring can be attributed to several factors. One factor which can account for both behaviors is that, in reflection-mode operation, the detector views a tilted projection of the sample surface. Accordingly, the reconstructed images of the sample will appear vertically compressed by a factor related to the incidence angle, which in this case is $\sin(\ang{45})$. Additionally, since our sample is a three-dimensional structure, the thickness of the the pattern set becomes more apparent as the sample surface tilts away from parallel orientation with the detector. The lengths of the line-like features (including the Siemens star spokes) are large (ca. $\SI{1}{\micro\metre}$) in comparison to its thickness (ca. $\SI{32}{\nano\metre}$), which leads to minor or unnoticeable changes to their appearances at this angle. However, the widths for the inner Siemens star spokes and several of the line patterns are more comparable to the pattern thickness, leading to a more noticeable change in their appearance. 

Another potential source of anisotropic resolution may be attributed to the X-ray interaction with the multilayer substrate itself. In focused-beam transmission ptychography measurements, the measured bright field produced by a uniformly-illuminated zone plate will appear as a annulus with largely uniform intensity. In contrast, the majority of the Bragg diffracted flux produced by the multilayer is concentrated within a narrow region of $2\theta$ on the detector, seen as a prominent ``band" on the annulus (i.e., the CCD image in Fig.~\ref{fig:instrument} and supplementary Fig. S1) which represents the rocking curve ($\theta$ scan) of the multilayer. In other words, the multilayer acts as an anisotropic angular filter which causes the illuminated region of the sample to see the filtered beam twice (i.e., entering into and exiting out of the multilayer). In contrast, the beam angles filtered out by the multilayer (i.e., outside of the rocking curve) will effectively interact with the sample only during incidence, thus reducing the amount of scattering those angles contribute to the measured coherent diffraction pattern. An additional effect may arise from the penetration depth of the X-ray beam into the multilayer, producing reflected beams that could emerge tens of nanometers downstream from their entry point. A combination of all these effects likely contributes to the anisotropic resolution in our reconstructed images. While these behaviors may limit the use of multilayered substrates as Bragg mirrors, they are not anticipated for other types samples. 

% /// THE ORIGINAL DISCUSSION ABOUT RESOLUTION BEFORE DAYNE MERGED THE PARAGRAPHS
Finally, we quantitatively assess our spatial resolution using the barcode patterns. In Fig.~\ref{fig:resolution}~(f) line-cuts are shown along the horizontal and vertical direction of the reconstructed barcode patterns. The features for the horizontal oriented bar pattern are clearly much sharper with more than twice the contrast. We used the two-dimensional Fourier ring correlation (FRC) analysis, as shown in Fig.~\ref{fig:resolution} (g) to estimate the resolution. This metric measures the achieved resolution based on the consistency between two independent reconstructions, split using pseudo-random sampling. Using the $1/2$ and $1/7$ criteria, the spatial resolution was evaluated in one configuration, yielding full-pitch resolutions of ca. $\SI{45}{\nano\metre}$ and ca. $\SI{30}{\nano\metre}$, respectively. Two independent reconstructions have been used in the probe view. It is worth mentioning, that FRC has notable limitations, including its sensitivity to image processing artifacts, and tendency to provide a single global resolution metric that may not reflect local variations in image quality across the field of view. Although line-cuts and visual inspection reveal poorer resolution along the beam direction, the Fourier ring correlation does not exhibit a commensurate loss, indicating that the observed anisotropy is governed by the measurement geometry rather than limitations of the reconstruction.

\section{Conclusion}
We have developed and demonstrated a scanning reflection soft X-ray ptychography microscope capable of ca. $\SI{45}{\nano\metre}$ spatial resolution for the first time. This method alleviates long-standing sample preparation constraints inherent to transmission microscopes and enables nanoscale investigation of specimens which could not previously be imaged with soft x-rays. A multilayer substrate was used to enhance reflectivity in a right-angle geometry. Benchmarking with a nano-structured test pattern demonstrated the highest spatial resolution for a reflection-mode instrument. Future extensions towards time resolution—including single-exposure full-frame measurements with randomized zone plates \cite{levitan2020}, coherent imaging modalities, and dichroic contrast hold promise for the comprehensive study of functional materials in situ with applied currents, fields or temperatures. Looking ahead, this reflection imaging approach is well positioned to benefit from ongoing synchrotron light-source upgrades, x-ray free electron laser developments that deliver much higher levels of coherent photon flux, enabling further gains in spatial and temporal resolution. It also opens new avenues for probing a wide variety of materials which have so far remained incompatible to  transmission-based microscopes.

\begin{backmatter}
\bmsection{Funding}
The LBNL LDRD Program FY24/25 supported salary of Damian Guenzing and the technical developments in this work. The research used the resources of the Advanced Light Source and the Molecular Foundry, which are both supported by the Office of Science, Office of Basic Energy Sciences, of the U.S. Department of Energy under Contract No. DE-AC02-05CH11231. Dayne Sasaki was supported under the ALS Postdoctoral fellowship (2024-25) and the Department of Energy, Office of Science, Office of Basic Energy Sciences, under Award Number DE-SC0021939. Abraham L. Levitan was supported by H2020 Marie Skłodowska-Curie Actions (884104 (PSI-FELLOW-III-3i)).                                                 

\bmsection{Acknowledgment} We thank Stephan Hruszkewycz and Yue Cao from Argonne National Laboratory for fruitful discussions. We also thank Emma Bernard for performing X-ray reflectometry measurements of the multilayer sample.

\bmsection{Disclosures}
\noindent The authors declare no conflicts of interest.

\bmsection{Data Availability Statement}
Data underlying the results presented in this paper are publicly available at \cite{gunzing_2025_dryad}.

% no supplemental for this now, since we only have it for the srxm at the moment shich is not really the core of that paper
%\bmsection{Supplemental document}
%See Supplement 1 for supporting content.

\end{backmatter}

% \cite{Zhang:14,OPTICA,FORSTER2007,Dean2006,testthesis,Yelin:03,Masajada:13,codeexample}.

% === SUPPLEMENTARY MATERIAL STARTS HERE ===
\setcounter{section}{0} % Reset section numbering
\renewcommand{\thesection}{S\arabic{section}} % Use S1, S2, ...
\setcounter{figure}{0} % Reset figure numbering
\renewcommand{\thefigure}{S\arabic{figure}} % Use S1, S2, ...
\setcounter{table}{0} % Reset table numbering
\renewcommand{\thetable}{S\arabic{table}} % Use S1, S2, ...
\newpage
\title{Supplementary Information for 'Soft X-ray Reflection Ptychography'}

\renewcommand{\thefigure}{S\arabic{figure}}
\renewcommand{\thetable}{S\arabic{table}}

\section{Sample Fabrication}

\subsection{Multilayer Substrate}
\label{sec:MLsubstrate}
\begin{figure}[ht]
    \centering 
    \includegraphics[width=0.75\textwidth]{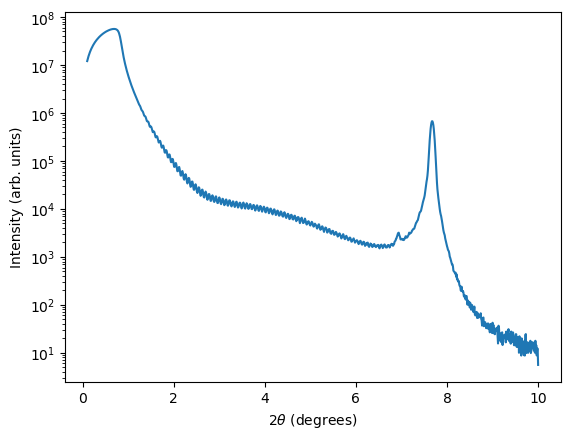} 
    \caption{Cu {$K_\alpha$} X-ray reflectivity measurement of the [Si/W] multilayer Bragg reflector used in this study.}
    \label{fig:SI_XRR}
\end{figure}

A multilayer Bragg reflector was used to maximize the specularly-reflected X-ray flux at the right-angle scattering geometry near a photon energy of $\SI{707}{\electronvolt}$. The multilayer consisted of 100 repeats of the [Si/W] bilayer unit deposited on a Si substrate. We used lab-based X-ray reflectivity measurements with a {Cu {$K_\alpha$} source to confirm the structure. The thickness of the [Si/W] bilayer unit was determined to be \SI{1.2}{\nano\metre} based on the location of the multilayer Bragg peak ($2\theta\approx\ang{7.67}$). The peak sitting to the left of the Bragg peak ($2\theta\approx\ang{6.93}$) is due to the incident X-ray beam containing Cu {$K_\beta$} radiation. Compared to a bare Si substrate, this multilayer substrate increased the reflectivity signal by approximately three orders of magnitude.

\subsection{Test Pattern}

A test pattern comprising a Siemens star and bar codes was fabricated on the multilayer substrate using electron-beam lithography followed by electron-beam evaporation. The pattern consisted of a \SI{3}{\nano\metre} Ti adhesion layer and a \SI{32}{\nano\metre} Au layer. The test pattern featured the following geometrical parameters:

\begin{itemize}
    \item \textbf{Bar codes:} Fixed length of \SI{1}{\micro\metre} with widths ranging from \SIrange{15}{200}{\nano\metre}. Vertically and horizontally oriented bar code patterns were identical, differing only by a \ang{90} rotation.
    \item \textbf{Siemens star:} Spoke tip widths of approximately \SI{40}{\nano\metre} and outer end widths of approximately \SI{250}{\nano\metre}. Both inner and outer spoke lengths were approximately \SI{1.2}{\micro\metre} with a spoke separation of approximately \SI{120}{\nano\metre}.
\end{itemize}

\section{Example Coherent Diffraction Pattern}
A representative coherent diffraction pattern from a single scan position is shown in Figure \ref{fig:SIexamplepattern} where the axes units represent the \SI{13.5}{\micro\metre} pixels and the intensity is in raw detector counts.

\begin{figure}[ht]
    \centering 
    \includegraphics[width=0.990\textwidth]{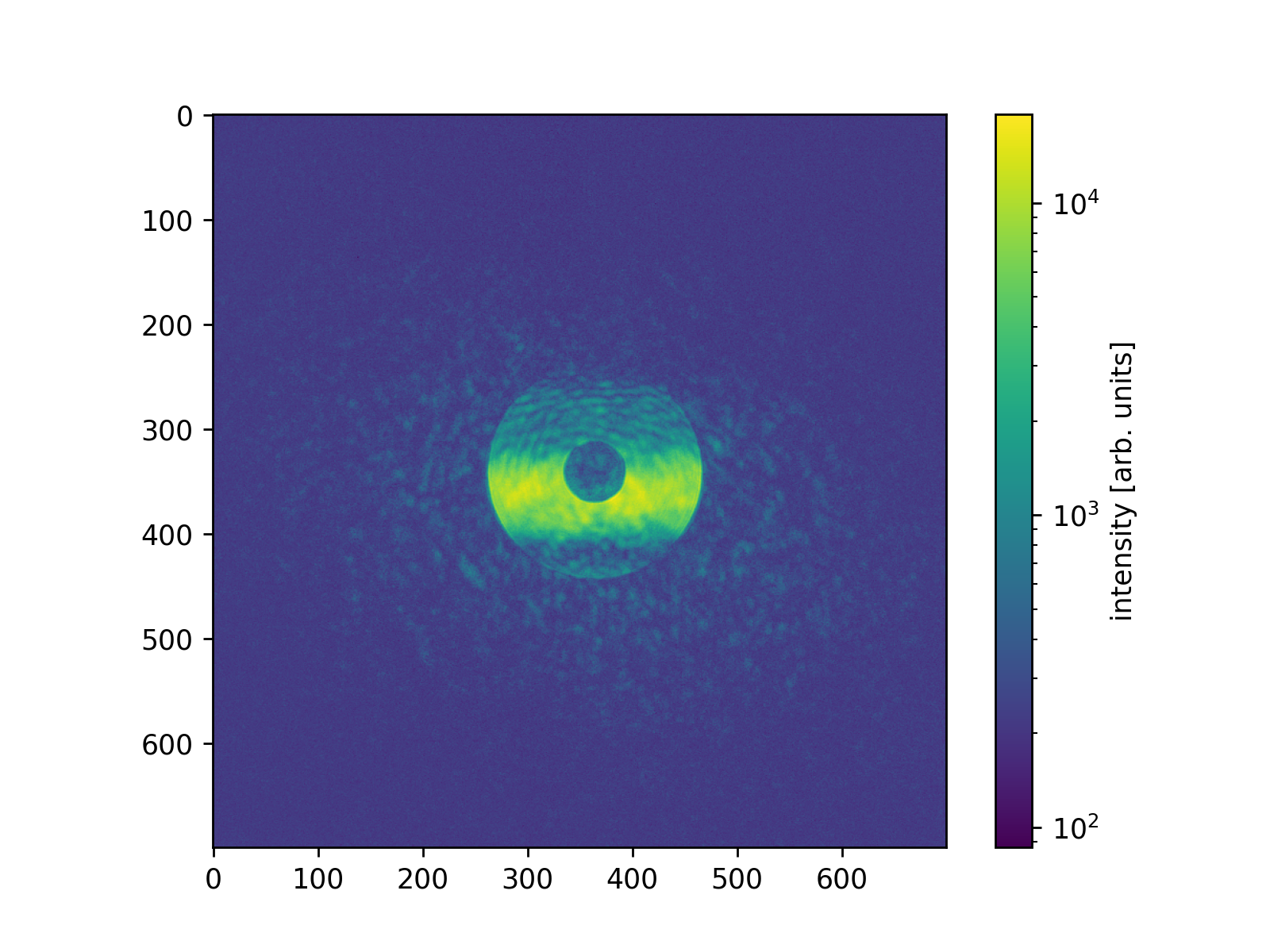} 
    \caption{Example of one of the collected diffraction patterns used for the reconstruction. Speckles are clearly visible around the main donut pattern representing the zoneplate illumination. A brighter band exists due to the multilayer used as discussed in the main text.}
    \label{fig:SIexamplepattern}
\end{figure}

\section{Ptychographic Data Acquisition}

The ptychographic dataset was acquired by collecting coherent diffraction patterns with a charge coupled device (CCD) as the zone plate illumination is rastered across the sample surface. For reflection geometry measurements, motion along the $y$-direction required a tandem motion along the $z$-direction to maintain a constant beam size across the tilted sample surface. The required $z$ motion as a function of $y$ displacement was determined trigonometrically from the sample surface slope relative to the $z$-axis.

The slope angle was experimentally determined by moving the beam along the $y$-direction by approximately \SI{50}{\micro\metre} to an adjacent pattern and refocusing. The slope angle was then calculated from the $z$ and $y$ positions of the two focused positions. Slight off mounting pushed the specular reflection energy to $\sim$\SI{760}{\electronvolt}. 

The ptychographic scan parameters were as follows:

\begin{table}[ht]
\centering
\begin{tabular}{ll}
\toprule
\textbf{Parameter} & \textbf{Value} \\
\midrule
Scan area & $\SI{10}{\micro\metre} \times \SI{10}{\micro\metre}$ \\
Grid size & $50 \times 50$ positions \\
Step size & \SI{200}{\nano\metre} \\
Acquisition time per position & \SI{1500}{\milli\second} \\
Zone plate--sample distance & \SI{10.7}{\milli\metre} \\
Beam spot diameter & $\sim$\SI{1.5}{\micro\metre} \\
Photon energy & $\sim$\SI{760}{\electronvolt} \\
Incidence angle & $\sim$\ang{45} \\
\bottomrule
\end{tabular}
\caption{Ptychographic scan parameters.}
\label{tab:scan_params}
\end{table}

Raw detector counts were converted to reflected intensity by subtracting the dark current measured before each scan.

\section{Ptychographic Reconstruction}

The reflection ptychography dataset can be reconstructed using the \textit{FancyPtycho} model implemented in the \textit{CDTools} ptychography package~\cite{cdtools}. Data and reconstruction script can be found in reference~\cite{gunzing_2025_dryad}

The reconstruction utilized the PyTorch-based Adam (Adaptive Moment Estimation) optimizer. Optimal reconstruction quality was achieved using one probe mode. The learning rate was decreased by a factor of 10 every 100 iterations to allow finer adjustments and more stable convergence as training progressed.

\section{Resolution Analysis}

Spatial resolution was assessed using Fourier Ring Correlation (FRC) analysis~\cite{vanheel2005}. The FRC method calculates frequency-dependent correlations between image pairs and determines resolution at the threshold where correlation drops below specified values. The analysis extracts lines based on stripe orientation, transforms them to frequency space, and computes correlations either directly or with frequency binning~\cite{shapiro2020}.

The spatial resolution was quantified using the $1/2$ and $1/7$ threshold criteria:

\begin{table}[ht]
\centering
\begin{tabular}{ll}
\toprule
\textbf{Criterion} & \textbf{Resolution} \\
\midrule
$1/2$ & $\sim$\SI{45}{\nano\metre} \\
$1/7$ & $\sim$\SI{30}{\nano\metre} \\
\bottomrule
\end{tabular}
\caption{Spatial resolution determined by FRC analysis.}
\label{tab:resolution}
\end{table}

\section{Transmission Simulation}
\label{sec:simulation}
The simulation discussed here can be performed using the script and layout file found in reference~\cite{gunzing_2025_dryad}. The expected amplitude image of the test pattern was simulated by calculating the X-ray path lengths through the Au patterns in the \ang{45} incidence geometry. For simplicity, we treat the multilayer substrate as an ideal mirror. Through this assumption, we rewrite this problem as a transmission geometry measurement through a three-dimensional model of our sample with double the thickness of gold used. The transmitted intensity was then calculated as:
\begin{equation}
I = I_0 \exp(-\mu \rho d) = I_0 \exp(-\alpha d)
\end{equation}
where $\mu$ is the mass absorption coefficient, $\rho$ is the material density, $\alpha$ is the attenuation length, and $d$ is the effective material thickness. The attenuation length of Au at \SI{760}{\electronvolt} was obtained from the CXRO X-ray Database as \SI{629.74}{\nano\metre}~\cite{HENKE1993181}.

%%%%%%%%%% using BibTeX:
%\bibliographystyle{plain}
\bibliography{betterreference}

\end{document}